\RequirePackage{fix-cm}
\documentclass[smallextended]{svjour3wide} 
\smartqed  
\usepackage{graphicx}
\usepackage{amsmath}
\usepackage{amssymb}
\usepackage{mathrsfs}
\usepackage{color}
\usepackage{bm}
\newcommand{\ave}[1]{\ensuremath{\left\langle#1\right\rangle}}
\newcommand{\aves}[1]{\ensuremath{\langle#1\rangle}}
\newcommand{\e}{\mathrm{e}}

\journalname{Journal of Statistical Physics}
\begin{document}

\title{Nonequilibrium Statistical Mechanics for Adiabatic Piston Problem}

\author{Masato Itami \and Shin-ichi Sasa}

\institute{M. Itami \and S. Sasa \at
              Department of Physics, Kyoto University, Kyoto 606-8502, Japan \\
              \email{itami@scphys.kyoto-u.ac.jp, sasa@scphys.kyoto-u.ac.jp}
}

\date{Received: date / Accepted: date}

\maketitle

\begin{abstract}
We consider the dynamics of a freely movable wall of mass $M$ with one degree of freedom that separates a long tube into two regions, each of which is filled with rarefied gas particles of mass $m$.
The gases are initially prepared at equal pressure but different temperatures, and we assume that the pressure and temperature of gas particles before colliding with the wall are kept constant over time in each region.
We elucidate the energetics of the setup on the basis of the local detailed balance condition, and then derive the expression for the heat transferred from each gas to the wall.
Furthermore, by using the condition, we obtain the linear response formula for the steady velocity of the wall and steady energy flux through the wall.
By using perturbation expansion in a small parameter $\epsilon\equiv\sqrt{m/M}$, we calculate the steady velocity up to order $\epsilon$.
\end{abstract}

\section{Introduction} \label{sec:intro}


Nonequilibrium transport phenomena in many cases have been investigated based on the linear response theory or the Onsager theory \cite{Groot-Mazur}.
Although it had been difficult to obtain useful relations for fluctuations beyond the linear response regime, non-trivial relations that are generally valid far from equilibrium, including the fluctuation theorem \cite{CrooksPRE1,CrooksPRE2,Evans-Cohen-Morriss,Gallavotti-Cohen,JarzynskiJSP,Kurchan,Lebowitz-Spohn,Maes,SeifertPRL} and the Jarzynski equality \cite{JarzynskiPRL}, were developed for the last two decades as a result of the time-reversal symmetry of microscopic mechanics.
Thanks to such universal relations, we can easily derive the well-known relations, such as the second law of thermodynamics, the McLennan ensembles, the Green--Kubo relations, and the Kawasaki nonlinear response relation \cite{CrooksPRE2,Hayashi-Sasa}.
Moreover, the newly discovered universal relations were confirmed by the laboratory experiments \cite{CollinETAL,LiphardtETAL,WangETAL} using small systems which are strongly influenced by fluctuations in their environment.
It should be noted that the universal relations were also utilized to estimate the rotary torque of $\mathrm{F}_1$-ATPase \cite{HayashiETAL}.


One of the important problems in small systems is to provide an energetic interpretation of phenomena.
In macroscopic systems, the thermodynamics is established with operationally identifying work as the energy transferred to a system accompanied with macroscopic volume change of the system caused by a macroscopic force and heat as the other energy transferred to the system through microscopic degrees of freedom.
On the other hand, although we can consider the energy transferred to a small system, it is still unsolved in the small system how to decompose the transferred energy into work and heat so as to be consistent with the results of the thermodynamics.
Sekimoto provided a reasonable definition of heat in Langevin systems \cite{SekimotoJPSJ,SekimotoBOOK}, while there are other stochastic systems where the energetics is still not fully understood.


In this paper, we consider the energetics of the following adiabatic piston problem \cite{Callen,Chernov-Lebowitz-Sinai,Feynman,Gruber-Lesne,Lieb}.
A freely movable wall of mass $M$ with one degree of freedom separates a long tube into two regions, each of which is filled with rarefied gas particles of mass $m$.
The wall is assumed to be thermally insulating and frictionless.
It is also assumed that $\epsilon \equiv \sqrt{m/M}$ is a small parameter, which controls the amount of energy transferred through the wall.
It should be noted that we need to work with a very small system for observing the motion of the wall in a laboratory experiment because $\epsilon$ in macroscopic systems is too small (less than $10^{-10}$).
If the wall were fixed, the energy could not be transferred through the wall.
Such a wall is referred to as ``adiabatic'' in the thermodynamic sense.
However, we note that the wall is not strictly adiabatic because the energy is transferred from the hot side to the cold side through the fluctuation of the wall, which is sometimes pointed out in the previous papers \cite{Kestemont-VandenBroeck-Mansour,Munakata-Ogawa}.
Thus, the wall can be regarded as a ``Brownian'' wall \cite{Van-Kawai-Meurs}.
The gases in the left and right regions are initially prepared at the same pressure $p$ but different temperatures $T_{\mathrm{L}}$ and $T_{\mathrm{R}}$, respectively.
Each gas is well approximated by an ideal gas, and it is also assumed that the pressure and temperature of gas particles before colliding with the wall are kept constant over time in each region.
In this case, the standard hydrodynamic equations suggest that the wall does not move due to the equal pressure.
However, perturbation methods for kinetic equations and molecular dynamics simulations \cite{Gruber-Frachebourg,Gruber-Pache-Lesne,Gruber-Piasecki,Kestemont-VandenBroeck-Mansour,Plyukhin-Schofield} reveal that the wall moves towards the hot side owing to the energy transfer from the hot side to the cold side  through the fluctuation of the wall.
Recently, a phenomenological mechanism for the emergence of the motion from the cross-coupling between momentum and heat flux has been proposed in Refs.~\cite{Fruleux-Kawai-Sekimoto,Kawai-Fruleux-Sekimoto}.


The local detailed balance condition can be helpful in defining heat in small systems where heat is not identified yet.
The local detailed balance condition states that when the system in contact with a single heat bath obeys the canonical distribution at the temperature of the heat bath, the ratio of probability density of the forward path and of the backward path is quantitatively related to the entropy production in the heat baths.
The local detailed balance condition holds in many systems including Hamiltonian systems \cite{JarzynskiJSP} and Langevin systems \cite{SeifertPRL}.
Thus, by using a model with the local detailed balance condition, we can define heat in the model.
Furthermore, since the local detailed balance condition immediately leads to most of the non-trivial relations that are generally valid far from equilibrium \cite{CrooksPRE2,SeifertRPP}, it plays a fundamental role in analyzing nonequilibrium systems.
Nevertheless, it should be noted that the local detailed balance condition is not obviously valid because the entropy production depends on a level of description \cite{Kawaguchi-Nakayama}.

In this paper, we provide a model for the adiabatic piston problem by using a continuous-time Markov jump process.
By correctly calculating the local detailed balance condition, we clarify that the entropy production depends on waiting times between jumps.
Furthermore, we provide the definition of heat in our model, and then elucidate the energetics.


This paper is organized as follows.
In Sect.~\ref{sec:model}, we explain our model.
In Sect.~\ref{sec:LDB}, we elucidate the energetics of our model on the basis of the local detailed balance condition.
In Sect.~\ref{sec:FT}, we derive several types of fluctuation theorems and the formal expression of the steady-state distribution.
In Sect.~\ref{sec:LRT}, we first show the Onsager theory for the adiabatic piston problem, and after that we derive the linear response formula.
We finally calculate one of the time-correlation functions explicitly, and derive the steady velocity of the wall up to order $\epsilon$.
The final section is devoted to a brief summary and remarks.
In Appendix, we confirm the validity of our model on the basis of Hamiltonian systems.
Throughout this paper, $\beta$ represents the inverse temperature and the Boltzmann constant $k_{\mathrm{B}}$ is set to unity.
The subscripts or superscripts $\mathrm{L}$ and $\mathrm{R}$ represent quantities on the left and right side, respectively.

\section{Model} \label{sec:model}

\subsection{Setup} \label{sec:setup}

\begin{figure}
 \centering
 \includegraphics[width=0.85\linewidth]{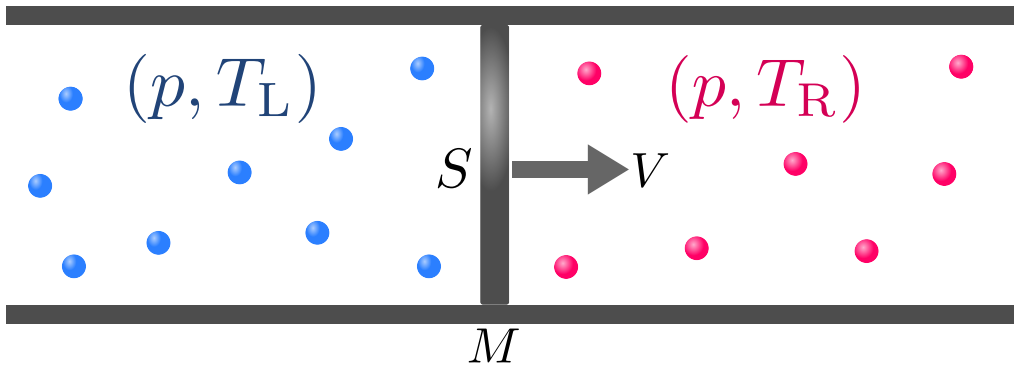}
 \caption{(color online) Schematic illustration of our model. A wall of mass $M$ with velocity $V$ separates a infinitely long tube of cross-sectional area $S$ into two regions, each of which is filled with rarefied gas particles of mass $m$. The gases in the left and right regions are initially prepared at equal pressure $p$ but different temperatures $T_{\mathrm{L}}$ and $T_{\mathrm{R}}$, respectively.}
 \label{fig:1}
\end{figure}

We introduce a model for studying the adiabatic piston problem.
A schematic illustration is shown in Fig.~\ref{fig:1}.
First, we provide a mechanical description of the wall of mass $M$.
We take the $x$-axis along the axial direction of an infinitely long tube of cross-sectional area $S$, and assume that the wall moves without friction along $x$-axis.
We denote by $V$ the velocity of the wall, which is the only degree of freedom of the wall.
When discussing time evolution of $V$, we denote by $V(t)$ its value at time $t$, and by $\hat{V}=(V(t))_{t\in [0,\tau]}$ its path during the time interval $[0,\tau]$.


Next, we provide an effective description of rarefied gas particles of mass $m$.
The gases in the left and right regions separated by the wall are initially prepared at equal pressure $p$ but different temperatures $T_{\mathrm{L}}$ and $T_{\mathrm{R}}$, respectively.
In the following, we focus on the gas on the left side; the gas on the right side can be described similarly.
We study a rarefied gas such that the characteristic time of the dissipation process inside each gas is much longer than the time during which we observe the steady state motion of the wall.
Therefore, gas particles that have yet to collide with the wall are in equilibrium at the temperature $T_{\mathrm{L}}$, the pressure $p$, and the number density $n_{\mathrm{L}}=p\beta_{\mathrm{L}}$.
We also assume that the gas particles elastically and instantaneously collide with the wall only once.
Furthermore, for simplicity, we assume that the surface of the wall is perpendicular to the $x$-axis, so that we consider only the $x$-component of the velocity of each gas particle.


For this setup, the interaction between the wall and the gas on the left side can be described by random collisions with the collision rate $\lambda_{\mathrm{L}}(v,V)$ for the gas particle velocity $v$ and the wall velocity $V$.
The collision rate is explicitly written as
\begin{equation}
 \lambda_{\mathrm{L}}(v,V) = n_{\mathrm{L}}S (v-V)\theta (v-V) f_{\mathrm{eq}}^{\mathrm{L}}(v),
  \label{eq:lambda_L}
\end{equation}
where $f_{\mathrm{eq}}^{\mathrm{L}}(v) =\sqrt{\beta_{\mathrm{L}} m/2\pi} \exp ( -\beta_{\mathrm{L}} mv^2 /2 )$ is the Maxwell--Boltzmann distribution and $\theta$ represents the Heaviside step function.
Similarly, the collision rate of the gas on the right side is given by
\begin{equation}
 \lambda_{\mathrm{R}}(v,V) = n_{\mathrm{R}}S(V-v) \theta (V-v)f_{\mathrm{eq}}^{\mathrm{R}}(v).
  \label{eq:lambda_R}
\end{equation}
By these effective descriptions of the gases, our model becomes a continuous-time Markov jump process.

\subsection{Time Evolution Equations} \label{sec:Time_evolution_equation}


We explicitly write an equation of motion of the wall.
When, due to collision, the velocities of a gas particle and the wall change from $v$ to $v'$ and from $V$ to $V'$, respectively, the laws of the conservation of energy and momentum are written as
\begin{align}
 v - V &= - v' + V',
 \label{eq:coefficient_restitution}\\
 mv+MV &= mv' + MV'.
 \label{eq:conservation_momentum}
\end{align}
Then, the impulse of the collision is given by
\begin{align}
 I(v,V) &=MV'-MV
 \notag
 \\
 &=\frac{2mM}{m+M}(v-V).
  \label{eq:impulse}
\end{align}
The $i$-th collision time of a gas particle and the wall at the left side is determined according to the Poisson process with the rate function, $\int dv\; \lambda_{\mathrm{L}}(v,V)$.
Suppose that a gas particle in the left side with a velocity $v_{i}^{{\mathrm{L}}}$ collides with the wall at $t=t_{i}^{\mathrm{L}}$.
The equation of motion of the wall is
\begin{equation}
M\frac{\mathrm{d} V}{\mathrm{d} t} = F_{{\mathrm{L}}} + F_{\mathrm{R}},
\label{eq:EOM_V}
\end{equation}
with
\begin{equation}
F_{{\mathrm{L}}} = \sum_{i} I\left( v_{i}^{{\mathrm{L}}}, \tilde{V}\right)
\delta\left( t-t_{i}^{{\mathrm{L}}}\right) ,
\label{eq:force_L}
\end{equation}
where $F_{\mathrm{L}}$ is the force exerted by the elastic collisions of the gas particles on the left side and $\tilde{V}(t)\equiv \lim_{t' \nearrow t} V(t')$ the velocity just before the collision when $t=t_{i}^{{\mathrm{L}}}$.
$F_{\mathrm{R}}$ is determined as well.
(\ref{eq:EOM_V}) describes the time evolution of $V(t)$ (or the path of $V(t)$).


Then, we derive a time evolution equation of the velocity distribution function at time $t$, $P(V,t)$, based on the equation of motion (\ref{eq:EOM_V}).
By using (\ref{eq:lambda_L}), (\ref{eq:lambda_R}), (\ref{eq:coefficient_restitution}), and (\ref{eq:conservation_momentum}), we obtain the following transition rate density from a state $V$ to another state $V'$:
\begin{equation}
 \omega\left( V\to V' \right) = \lambda (v,V) \frac{\mathrm{d}v}{\mathrm{d}V'},
  \label{eq:transition_rate}
\end{equation}
where
\begin{equation}
 \lambda (v,V) \equiv \lambda_{\mathrm{L}} (v,V) + \lambda_{\mathrm{R}} (v,V),
\end{equation}
and
\begin{equation}
 v=\frac{(M+m)V'-(M-m)V}{2m}.
\end{equation}
Moreover, by using (\ref{eq:transition_rate}), the escape rate is given by
\begin{align}
 \kappa(V) &= \int \mathrm{d}V' \; \omega \left( V \to V' \right)
 \notag
 \\
 &= \int \mathrm{d}v\; \lambda (v,V).
  \label{eq:escape_rate}
\end{align}
In terms of the transition rate density and the escape rate, we express the time evolution equation of $P(V,t)$ as
\begin{equation}
\frac{\partial P(V,t)}{\partial t}=\int \mathrm{d}V''\; \omega (V''\to V) P(V'',t) - \kappa(V) P(V,t).
\label{eq:mastereq_omega}
\end{equation}
By using (\ref{eq:transition_rate}), we can rewrite (\ref{eq:mastereq_omega}) as 
\begin{equation}
\frac{\partial P(V,t)}{\partial t}=\int \mathrm{d}v\; \lambda (v,V'') \frac{\mathrm{d} V''}{\mathrm{d} V} P(V'',t) - \kappa(V) P(V,t),
\label{eq:mastereq_lambda}
\end{equation}
where
\begin{equation}
 V''=V-\frac{2m}{M-m}(v-V).
\end{equation}
(\ref{eq:mastereq_lambda}) is called the master-Boltzmann equation.
It should be noted that, when $T_{\mathrm{L}}=T_{\mathrm{R}}=T$,  (\ref{eq:lambda_L}), (\ref{eq:lambda_R}), (\ref{eq:coefficient_restitution}), (\ref{eq:conservation_momentum}), and (\ref{eq:transition_rate}) lead to the detailed balance condition:
\begin{equation}
 P_{\mathrm{eq}}(V)\omega (V\to V'') = P_{\mathrm{eq}}(-V'')\omega (-V''\to -V),
  \label{eq:detailed_balance_condition}
\end{equation}
where we denote the Maxwell--Boltzmann distribution by
\begin{equation}
P_{\mathrm{eq}}(V) = \sqrt{\frac{\beta M}{2\pi}}\e^{-\beta \frac{MV^2}{2}}.
\label{eq:eqilibrium_distribution}
\end{equation}
This supports the validity of our model in equilibrium.

\subsection{Notations} \label{sec:Notations}


For later convenience, we define physical quantities.
Given a path $\hat{V}$, we denote the total number of collisions as $n$, the time at the $i$-th collision as $t_i$, and the velocity after the $i$-th collision as $V_{i}$, where $t_0\equiv 0$, $t_{n+1}\equiv \tau$, and $V(0)\equiv V_0$.
We write the time reversal of $V$ and $\hat{V}$ as $V^{*} = -V$ and $\hat{V}^{\dagger} =(V^{*} (\tau - t))_{t\in [0,\tau]}$, respectively.
In the following, we denote the mean inverse temperature and the degree of nonequilibrium by $\beta \equiv (\beta_{\mathrm{L}}+\beta_{\mathrm{R}})/2$ and $\Delta \equiv (\beta_{\mathrm L}-\beta_{\mathrm R})/\beta$, respectively.
Throughout this paper, calligraphic fonts mean that its quantity depends on the path $\hat{V}$.

\section{Local Detailed Balance Condition} \label{sec:LDB}

\subsection{Naive Consideration} \label{sec:naive_consideration}

In order to elucidate the energetics of this model, we first calculate $\omega ( V\to V') / \omega ( V'^{*}\to V^{*})$ because it has been known that the ratio is related to the entropy production of the heat baths in many cases.
We consider the case where, due to collision, the velocities of a gas particle and the wall change from $v$ to $v'$ and from $V$ to $V'$, respectively.
Then, by using (\ref{eq:coefficient_restitution}) and (\ref{eq:conservation_momentum}), we can show that the velocity of the wall changes from $V'^{*}$ to $V^{*}$ when that of the bath particle changes from $v'^{*}$ to $v^{*}$.
Furthermore, we obtain
\begin{equation}
 V-v=\frac{m+M}{2m}(V-V') = V'^{*} - v'^{*},
 \label{velo_diff}
\end{equation}
which means that $V < v \Leftrightarrow V < V' \Leftrightarrow V'^{*} < v'^{*}$.
Thus, (\ref{eq:transition_rate}) leads to
\begin{align}
 \frac{\omega (V\to V')}{\omega (V'^{*}\to V^{*})} &= \frac{\lambda_{\mathrm{L}} (v,V)}{\lambda_{\mathrm{L}} (v'^{*},V'^{*})}\theta (v-V) + \frac{\lambda_{\mathrm{R}} (v,V)}{\lambda_{\mathrm{R}} (v'^{*},V'^{*})} \theta (V-v)
 \notag
 \\[5pt]
 & = \mathrm{e}^{-\beta_{\mathrm{L}}\left[ \frac{mv^2}{2}-\frac{m{v'}^2}{2}\right] \theta (v-V)-\beta_{\mathrm{R}}\left[ \frac{mv^2}{2}-\frac{m{v'}^2}{2}\right] \theta (V-v)}
 \notag
 \\[5pt]
 & = \mathrm{e}^{-\beta_{\mathrm{L}}\left[ \frac{MV'^2}{2}-\frac{MV^2}{2}\right] \theta (V'-V) -\beta_{\mathrm{R}}\left[ \frac{MV'^2}{2}-\frac{MV^2}{2} \right] \theta (V-V')}
 ,\label{eq:omega}
\end{align}
where we have used the conservation of kinetic energy in elastic collisions.
Therefore, given a path $\hat{V}$, we obtain
\begin{equation}
 \prod_{i=1}^{n} \frac{\omega ( V_{i-1}\to V_{i})}{\omega ( V_{i}^{*}\to V_{i-1}^{*})} = \mathrm{e}^{-\beta_{\mathrm{L}}\mathcal{K}_{\mathrm{L}}(\hat{V}) - \beta_{\mathrm{R}}\mathcal{K}_{\mathrm{R}}(\hat{V})},
  \label{eq:prod_omega}
\end{equation}
with
\begin{equation}
 \begin{split}
  \mathcal{K}_{\mathrm{L}}(\hat{V}) &\equiv \sum_{i=1}^{n} \left( \frac{MV_{i}^2}{2} - \frac{MV_{i-1}^2}{2} \right) \theta (V_{i} - V_{i-1}),\\
  \mathcal{K}_{\mathrm{R}}(\hat{V}) &\equiv \sum_{i=1}^{n} \left( \frac{MV_{i}^2}{2} - \frac{MV_{i-1}^2}{2} \right) \theta (V_{i-1}-V_{i}),
 \end{split}
 \label{eq:form_energy_transferred}
\end{equation}
where we denote the total increment of the kinetic energy of the wall by the collisions of the gas particles on the left and right side during the time interval $[0,\tau]$ by $\mathcal{K}_{\mathrm{L}}(\hat{V})$ and $\mathcal{K}_{\mathrm{R}}(\hat{V})$, respectively.
They satisfy
\begin{equation}
 \mathcal{K}_{\mathrm{L}}(\hat{V})+\mathcal{K}_{\mathrm{R}}(\hat{V})=\frac{MV_n^2}{2}-\frac{MV_0^2}{2}.
\end{equation}
Since collisions between the wall and each gas particle are elastic, $\mathcal{K}_{\mathrm{L}}(\hat{V})$ is equal to the decrease in the total kinetic energy of the gas particles on the left side.
Thus, in this model, $\mathcal{K}_{\mathrm{L}}(\hat{V})$ and $\mathcal{K}_{\mathrm{R}}(\hat{V})$ are the energy transferred from the left and right side to the wall during the time interval $[0,\tau ]$, respectively.

If the wall is fixed, which is a case considered in many examples, the energy transferred $\mathcal{K}_{\mathrm{L}}(\hat{V})$ may be interpreted as the heat transferred from the left side to the wall.
Indeed, it was assumed that the ratio of the transition rates $\omega (V\to V')$ and $\omega (V'^{*}\to V^{*})$ is equal to the exponential of the entropy production in the formal arguments \cite{Sagawa-Hayakawa}.
However, since the wall can move in the model under consideration, work is done by the gas particles on each side.
In order to obtain the proper entropy production, instead of the transition rate $\omega (V\to V')$, we have to precisely consider the probability density of the path $\hat{V}$ under the condition that $V_0$ is given.

\subsection{True Expression} \label{sec:LDBC}

The probability density of the path $\hat{V}$ for a given $V_0$ is expressed as
\begin{equation}
 \mathcal{P}_{\Delta}(\hat{V}\vert V_0) = \e^{-\kappa (V_{0}) \times (t_1 - t_0)} \prod_{i=1}^{n} \omega ( V_{i-1}\to V_{i})\; \e^{-\kappa (V_{i}) \times (t_{i+1}-t_{i})}.
 \label{eq:weight}
\end{equation}
By using a certain initial distribution, $P_{\mathrm{ini}}(V_0)$, the expectation of any path-dependent quantity $\mathcal{A}(\hat{V})$ over all paths is given by
\begin{equation}
 \ave{\mathcal{A}}_{\Delta} \equiv \int \mathcal{D}\hat{V} \; P_{\mathrm{ini}}(V_0) \mathcal{P}_{\Delta}(\hat{V}\vert V_0) \mathcal{A}(\hat{V}),
\end{equation}
where we denote the integral over all paths by
\begin{equation}
 \int \mathcal{D}\hat{V} \; \equiv \sum_{n=0}^{\infty}\; \int_{-\infty}^{\infty}dV_{n}\cdots \int_{-\infty}^{\infty}dV_{0} \int_{0}^{\tau} dt_{n} \int_0^{t_{n}}dt_{n-1}\cdots \int_{0}^{t_2}dt_1 ,
  \label{eq:integral_all_paths}
\end{equation}
which satisfies $\mathcal{D}\hat{V} = \mathcal{D}\hat{V}^{\dagger}$.
We also denote the probability density in equilibrium ($T_{\mathrm{L}}=T_{\mathrm{R}}$) by $\mathcal{P}_{0}(\hat{V}\vert V_0)$ and the expectation of $\mathcal{A}(\hat{V})$ in equilibrium by
\begin{equation}
 \ave{\mathcal{A}}_{0} \equiv \int \mathcal{D}\hat{V} \; P_{\mathrm{ini}}(V_0) \mathcal{P}_{0}(\hat{V}\vert V_0) \mathcal{A}(\hat{V}).
\end{equation}

Here, the escape rate $\kappa(V)$ is calculated as
\begin{align}
 \kappa(V) &= \int \mathrm{d}v \; \lambda (v,V)
 \notag
 \\[5pt]
 & = pS\beta_{\mathrm{L}} \int_{V}^{\infty} \mathrm{d}v\; (v-V) f_{\mathrm{eq}}^{\mathrm{L}}(v) + pS\beta_{\mathrm{R}} \int_{-\infty}^{V}\mathrm{d}v\; (V-v)f_{\mathrm{eq}}^{\mathrm{R}}(v).
\end{align}
This leads to
\begin{align}
 \kappa(V) - \kappa(V^*) &= pS\beta_{\mathrm{L}} \int_{-\infty}^{\infty}\mathrm{d}v\;(v-V)f_{\mathrm{eq}}^{\mathrm{L}}(v) + pS\beta_{\mathrm{R}} \int_{-\infty}^{\infty}\mathrm{d}v\; (V-v) f_{\mathrm{eq}}^{\mathrm{R}}(v)
 \notag
 \\[5pt]
 &= -(\beta_{\mathrm{L}}-\beta_{\mathrm{R}})pSV.
 \label{eq:kappa}
\end{align}
It should be noted that $\kappa (V)\neq \kappa (V^{*})$ when $\beta_{\mathrm{L}}\neq \beta_{\mathrm{R}}$.

Thus, by using (\ref{eq:prod_omega}), (\ref{eq:weight}), and (\ref{eq:kappa}), we obtain
\begin{align}
 \frac{\mathcal{P}_{\Delta}(\hat{V}\vert V_0)}{\mathcal{P}_{\Delta}(\hat{V}^{\dagger}\vert V_{n}^{*})}
 & = \prod_{i=1}^{n} \frac{\omega ( V_{i-1}\to V_{i})}{\omega ( V_{i}^{*}\to V_{i-1}^{*})} \prod_{i=0}^{n} \mathrm{e}^{-\left[ \kappa(V_{i})-\kappa(V_{i}^{*})\right] \times (t_{i+1}-t_{i})}
 \notag
 \\[5pt]
 & = \mathrm{e}^{-\beta_{\mathrm{L}}\left[ \mathcal{K}_{\mathrm{L}}(\hat{V}) - pS\mathcal{X}(\hat{V})\right] - \beta_{\mathrm{R}}\left[ \mathcal{K}_{\mathrm{R}}(\hat{V})+ pS\mathcal{X}(\hat{V})\right]},
 \label{eq:LDB}
\end{align}
where we have defined
\begin{equation}
 \mathcal{X}(\hat{V}) \equiv \sum_{i=0}^{n} V_{i} \; (t_{i+1} - t_{i}),
  \label{eq:def_X}
\end{equation}
which represents the displacement of the wall during the time interval $[0,\tau]$.
Here, we define
\begin{equation}
 \begin{split}
  \mathcal{Q}_{\mathrm{L}}(\hat{V}) &\equiv \mathcal{K}_{\mathrm{L}}(\hat{V})-pS\mathcal{X}(\hat{V}),\\
  \mathcal{Q}_{\mathrm{R}}(\hat{V}) &\equiv \mathcal{K}_{\mathrm{R}}(\hat{V})+pS\mathcal{X}(\hat{V}).
 \end{split}
 \label{eq:form_heat}
\end{equation}
If $\mathcal{Q}_{\mathrm{L}}$ and $\mathcal{Q}_{\mathrm{R}}$ are the heat transferred from the left and right side to the wall during the time interval $[0,\tau]$, the equality (\ref{eq:LDB}) is called the local detailed balance condition, the microscopically reversible condition \cite{CrooksPRE1}, or the detailed fluctuation theorem \cite{JarzynskiJSP}.
When the system in contact with a single heat bath obeys the canonical distribution at the temperature of the heat bath, we can expect that the local detailed balance condition is valid.
Indeed, these definitions of the heat transferred are reasonable, because the work done by the gas particles in the left side is equal to $pS\mathcal{X}(\hat{V})$ and (\ref{eq:form_heat}) corresponds to the first law of thermodynamics.
Once we derive the true expression of the local detailed balance condition, there is no difficulty of the understanding.
Nevertheless, we wish to emphasize that it is not easy to conjecture that the difference of the escape rates, $\kappa(V_{i})-\kappa(V_{i}^{*})$, contributes to the entropy production in a concrete physical model although it is known in general cases \cite{Maes-Netocny-Wynants}.
By studying a concrete example on the basis of the local detailed balance condition, we have reached the consistent decomposition of the energy transferred into the heat transferred and the work.

Furthermore, we define the heat transferred from right to left by
\begin{equation}
 \mathcal{Q}(\hat{V}) \equiv \frac{\mathcal{Q}_{\mathrm{R}}(\hat{V}) - \mathcal{Q}_{\mathrm{L}}(\hat{V})}{2}.
\end{equation}
Then, by using
\begin{align}
 \mathcal{Q}_{\mathrm{L}}(\hat{V})+\mathcal{Q}_{\mathrm{R}}(\hat{V}) &= \mathcal{K}_{\mathrm{L}}(\hat{V})+\mathcal{K}_{\mathrm{R}}(\hat{V})
 \notag
 \\
 &= \frac{MV_{n}^{2}}{2}-\frac{MV_{0}^{2}}{2},
\end{align}
we can rewrite the local detailed balance condition (\ref{eq:LDB}) as
\begin{equation}
\frac{P_{\mathrm{eq}}(V_0) \mathcal{P}_{\Delta}(\hat{V}\vert V_0)}{P_{\mathrm{eq}}(V_{n}^{*}) \mathcal{P}_{\Delta}(\hat{V}^{\dagger}\vert V_{n}^{*})} = \mathrm{e}^{ \Delta \beta \mathcal{Q}(\hat{V})}.
\label{eq:LDB2}
\end{equation}
This expression of the local detailed balance condition leads to several types of fluctuation theorems and the formal expression of the steady-state distribution, which are useful for easily deriving the well-known relations \cite{CrooksPRE1}.
It should be noted that the local detailed balance condition can also be derived from Hamiltonian systems, where the degrees of freedom of the gas particles are explicitly considered.
See Appendix for a detailed explanation.

\section{Fluctuation Theorem}\label{sec:FT}

In order to obtain concise expressions, we assume that the initial distribution of the velocity of the wall, $P_{\mathrm{ini}}(V_0)$, is given by the Maxwell--Boltzmann distribution, $P_{\mathrm{eq}}(V_0)$.
For any path-dependent quantity $\mathcal{A}(\hat{V})$, we define its time reversal by $\mathcal{A}^{\dagger}(\hat{V})\equiv \mathcal{A}(\hat{V}^{\dagger})$.

First, by using (\ref{eq:LDB2}), we obtain 
\begin{align}
 \ave{\mathcal{A}}_{\Delta} &= \int \mathcal{D}\hat{V} \; P_{\mathrm{eq}}(V_0) \mathcal{P}_{\Delta}(\hat{V}\vert V_0) \mathcal{A}(\hat{V})
 \notag
 \\[5pt]
 & = \int \mathcal{D}\hat{V}^{\dagger} \; P_{\mathrm{eq}}(V_{n}^{*}) \mathcal{P}_{\Delta}(\hat{V}^{\dagger}\vert V_{n}^{*}) \mathcal{A}^{\dagger}(\hat{V}^{\dagger}) \mathrm{e}^{\Delta \beta \mathcal{Q}^{\dagger}(\hat{V}^{\dagger})}
 \notag
 \\[5pt]
 & = \ave{\mathcal{A}^{\dagger} \mathrm{e}^{\Delta \beta \mathcal{Q}^{\dagger}}}_{\Delta} .
 \label{eq:useful_equality}
\end{align}
By setting $\mathcal{A}(\hat{V})\equiv 1$ and using $\mathcal{Q}^{\dagger}(\hat{V})=-\mathcal{Q}(\hat{V})$, we obtain the integral fluctuation theorem:
\begin{equation}
 \ave{\mathrm{e}^{-\Delta \beta \mathcal{Q}}}_{\Delta} = 1.
  \label{eq:Integral_FT}
\end{equation}
Jensen's inequality leads to
\begin{equation}
 \Delta \beta \aves{\mathcal{Q}}_{\Delta}\geq 0.
\end{equation}

Next, we derive the symmetry of the generating function \cite{Kurchan,Lebowitz-Spohn}.
We define the energy transferred from right to left by
\begin{equation}
 \mathcal{K}(\hat{V}) \equiv \frac{\mathcal{K}_{\mathrm{R}}(\hat{V})-\mathcal{K}_{\mathrm{L}}(\hat{V})}{2},
\end{equation}
and the scaled cumulant generating function by
\begin{equation}
 G(h_1, h_2) \equiv \lim_{\tau \to \infty} -\frac{1}{\tau} \log \ave{ \mathrm{e}^{-h_1\mathcal{K}-h_2 \mathcal{X}}}_{\Delta}.
  \label{eq:generating_func}
\end{equation}
By using (\ref{eq:useful_equality}) with $\mathcal{A}(\hat{V}) =\mathrm{e}^{-h_1 \mathcal{K}(\hat{V}) -h_2 \mathcal{X}(\hat{V})}$, $\mathcal{K}^{\dagger}(\hat{V})=-\mathcal{K}(\hat{V})$, and $\mathcal{X}^{\dagger}(\hat{V})=-\mathcal{X}(\hat{V})$, we obtain
\begin{equation}
 G(h_1,h_2) = G(\Delta\beta -h_1, \Delta\beta pS-h_2).
  \label{eq:symmetry_generating_func}
\end{equation}


Finally, we derive the steady-state distribution of the velocity of the wall.
We denote the path ensemble average of $\mathcal{A}(\hat{V})$ in equilibrium with the initial condition $V_0=V$ as
\begin{equation}
 \overline{\mathcal{A}}(V) \equiv \lim_{\tau \to \infty} \int \mathcal{D}\hat{V}\; \delta(V_0 - V) \mathcal{P}_{0}(\hat{V}\vert V_0) \mathcal{A}(\hat{V}).
  \label{eq:ave_eq_ini}
\end{equation}
The steady state distribution function $P_{\mathrm{st}}(V)$ is formally obtained as
\begin{equation}
 P_{\mathrm{st}}(V)=\lim_{\tau \to \infty} \ave{\delta (V(\tau)-V)}_{\Delta}.
  \label{eq:st_dist}
\end{equation}
Thus, by using (\ref{eq:useful_equality}) with $\mathcal{A}(\hat{V})=\delta (V(\tau)-V)$, $\delta (V_0^{*}-V)=\delta (V_0-V^{*})$, and space-reflection symmetry in equilibrium that leads to $\overline{\mathcal{Q}}(V^{*})=-\overline{\mathcal{Q}}(V) $, we obtain
\begin{align}
 P_{\mathrm{st}}(V) &= \lim_{\tau \to \infty} \ave{\delta (V_0-V^{*})\mathrm{e}^{-\Delta \beta \mathcal{Q}}}_{\Delta}
 \notag
 \\[5pt]
 & = P_{\mathrm{eq}}(V^{*}) \; \mathrm{e}^{-\Delta \beta \overline{\mathcal{Q}}(V^{*})+O(\Delta^2)}
 \notag
 \\[5pt]
 & = P_{\mathrm{eq}}(V) \; \mathrm{e}^{\Delta \beta \overline{\mathcal{Q}}(V) +O(\Delta^2)} ,
\label{eq:st_dist_LRR}
\end{align}
which is called the McLennan ensemble \cite{Hayashi-Sasa,Maes-Netocny,Mclennan,Zubarev}.

\section{Linear Response Theory} \label{sec:LRT}

\subsection{Onsager Theory for Adiabatic Piston Problem}\label{sec:Onsager_adiabatic_piston}

In this subsection, we denote the pressures of the gases on the left and right sides by $p_{\mathrm{L}}$ and $p_{\mathrm{R}}$, respectively.
In the following, we assume that the system settles to a unique nonequilibrium steady state when it evolves for a sufficiently long time.
Furthermore, we assume that thermodynamic forces, $\beta_{\mathrm{L}}-\beta_{\mathrm{R}}$ and $\beta_{\mathrm{L}} p_{\mathrm{L}} -\beta_{\mathrm{R}} p_{\mathrm{R}}$, are small compared to respective reference values.
We define by $J_{K}$ and $J_{V}$ the steady energy flux from right to left and the steady velocity of the wall in the linear response regime, respectively.
Within the linear response regime, Onsager's phenomenological equations relate the thermodynamic forces and fluxes as
\begin{equation}
 \begin{cases}
  J_{K} = \mathrm{L}_{11} (\beta_{\mathrm{L}}-\beta_{\mathrm{R}}) + \mathrm{L}_{12} (\beta_{\mathrm{L}} p_{\mathrm{L}} -\beta_{\mathrm{R}} p_{\mathrm{R}}),
  \\[5pt]
  J_{V} = \mathrm{L}_{21} (\beta_{\mathrm{L}}-\beta_{\mathrm{R}}) + \mathrm{L}_{22} (\beta_{\mathrm{L}} p_{\mathrm{L}} -\beta_{\mathrm{R}} p_{\mathrm{R}}),
 \end{cases}
 \label{eq:Onsager_energy}
\end{equation}
where $\mathrm{L}_{ij}$ are Onsager coefficients.
Onsager's reciprocity relation states that $\mathrm{L}_{12}=\mathrm{L}_{21}$.
Considering $\mathcal{Q} = \mathcal{K} + pS\mathcal{X}$, we define the steady heat flux from right to left in the linear response regime by
\begin{equation}
 J_{Q} \equiv J_{K} + \overline{p}J_{V},
\end{equation}
with
\begin{equation}
 \overline{p} \equiv \frac{p_{\mathrm{L}}+p_{\mathrm{R}}}{2}.
\end{equation}
Then, we obtain
\begin{equation}
 \begin{cases}
  J_{Q} = \widetilde{{\mathrm L}}_{11} (\beta_{\mathrm{L}}-\beta_{\mathrm{R}}) + \widetilde{{\mathrm L}}_{12} \beta (p_{\mathrm{L}}-p_{\mathrm{R}}),\\[5pt]
  J_{V} = \widetilde{{\mathrm L}}_{21} (\beta_{\mathrm{L}}-\beta_{\mathrm{R}}) + \widetilde{{\mathrm L}}_{22} \beta (p_{\mathrm L}-p_{\mathrm R}),
 \end{cases}
 \label{eq:Onsager_heat}
\end{equation}
with
\begin{equation} 
 \begin{pmatrix}
  \widetilde{\mathrm{L}}_{11} & \widetilde{\mathrm{L}}_{12}\\
  \widetilde{\mathrm{L}}_{21} & \widetilde{\mathrm{L}}_{22}
 \end{pmatrix}
 =
 \begin{pmatrix}
  \mathrm{L}_{11}+ \mathrm{L}_{12}\overline{p} + \mathrm{L}_{21}\overline{p} + \mathrm{L}_{22}\overline{p}^2 & \mathrm{L}_{12} + \mathrm{L}_{22} \overline{p}\\
  \mathrm{L}_{21}+ \mathrm{L}_{22}\overline{p} & \mathrm{L}_{22}
 \end{pmatrix},
 \label{eq:Onsager_coefficient_relation}
\end{equation}
where we have used
\begin{equation}
 \beta_{\mathrm{L}} p_{\mathrm{L}} -\beta_{\mathrm{R}} p_{\mathrm{R}} = \overline{p}(\beta_{\mathrm{L}}-\beta_{\mathrm{R}}) +\beta(p_{\mathrm{L}}-p_{\mathrm{R}}).
\end{equation}
In this form, the reciprocity relation, $\widetilde{{\mathrm L}}_{12}=\widetilde{{\mathrm L}}_{21}$, is also satisfied.
In order to calculate $J_{Q}$ and $J_{V}$, we express the Onsager coefficients in terms of the time correlation functions.
It should be noted that $p_{\mathrm{L}}=p_{\mathrm{R}}=p$ in our model.

\subsection{Linear Response Formula}\label{sec:LRformula}


By using (\ref{eq:useful_equality}) with $\mathcal{A}(\hat{V})=\mathcal{K}(\hat{V}) / (\tau S)$, we obtain
\begin{align}
 \lim_{\tau \to \infty} \ave{\frac{\mathcal{K}}{\tau S}}_{\Delta} &= \lim_{\tau \to \infty} \ave{-\frac{\mathcal{K}}{\tau S}\; \mathrm{e}^{-\Delta \beta \mathcal{Q}}}_{\Delta}
 \notag
 \\[5pt]
 & = \lim_{\tau \to \infty} \left[ -\ave{\frac{\mathcal{K}}{\tau S}}_{\Delta} + \frac{\Delta \beta}{\tau S} \ave{\mathcal{K}\mathcal{Q}}_{0} + O(\Delta^2) \right] .
 \label{eq:pre_LRformula_K}
\end{align}
This leads to
\begin{align}
 \lim_{\tau \to \infty} \ave{\frac{\mathcal{K}}{\tau S}}_{\Delta} &= \Delta \beta \lim_{\tau \to \infty} \frac{1}{2\tau S}\ave{\mathcal{K}\mathcal{Q}}_{0} + O(\Delta^2)
 \notag
 \\[5pt]
 & =\Delta \beta \lim_{\tau \to \infty} \frac{1}{2\tau S}\ave{\mathcal{K}\left[ \mathcal{K}+pS\mathcal{X}\right]}_{0} + O(\Delta^2).
 \label{eq:LRformula_K}
\end{align}
Similarly, by using (\ref{eq:useful_equality}) with $\mathcal{A}(\hat{V})=\mathcal{X}(\hat{V}) / \tau$, we obtain
\begin{align}
  \lim_{\tau \to \infty} \ave{\frac{\mathcal{X}}{\tau}}_{\Delta} &= \Delta \beta \lim_{\tau \to \infty} \frac{1}{2\tau}\ave{\mathcal{X}\mathcal{Q}}_{0} + O(\Delta^2)
 \notag
 \\[5pt]
 & =\Delta \beta \lim_{\tau \to \infty} \frac{1}{2\tau }\ave{\mathcal{X}\left[ \mathcal{K}+pS\mathcal{X}\right]}_{0} + O(\Delta^2).
 \label{eq:LRformula_V}
\end{align}
It should be noted that these universal relations, (\ref{eq:LRformula_K}) and (\ref{eq:LRformula_V}), hold for any $\epsilon=\sqrt{m/M}$.
Considering (\ref{eq:Onsager_energy}), (\ref{eq:LRformula_K}), and (\ref{eq:LRformula_V}), we obtain
\begin{equation}
  \begin{pmatrix}
  \mathrm{L}_{11} & \mathrm{L}_{12}\\
  \mathrm{L}_{21} & \mathrm{L}_{22}
 \end{pmatrix}
 =
 \begin{pmatrix}
  \displaystyle \lim_{\tau \to \infty} \frac{1}{2\tau S}\ave{\mathcal{K}\mathcal{K}}_{0} & \displaystyle \lim_{\tau \to \infty} \frac{1}{2\tau}\ave{\mathcal{K}\mathcal{X}}_{0}\\[8pt]
  \displaystyle \lim_{\tau \to \infty} \frac{1}{2\tau}\ave{\mathcal{X}\mathcal{K}}_{0} & \displaystyle \lim_{\tau \to \infty} \frac{S}{2\tau}\ave{\mathcal{X}\mathcal{X}}_{0}
 \end{pmatrix}.
 \label{eq:Onsager_coefficient_energy_expression}
\end{equation}
Furthermore, (\ref{eq:Onsager_coefficient_relation}) and (\ref{eq:Onsager_coefficient_energy_expression}) lead to
\begin{equation}
 \begin{pmatrix}
  \widetilde{\mathrm{L}}_{11} & \widetilde{\mathrm{L}}_{12}\\
  \widetilde{\mathrm{L}}_{21} & \widetilde{\mathrm{L}}_{22}
 \end{pmatrix}
 =
 \begin{pmatrix}
  \displaystyle \lim_{\tau \to \infty} \frac{1}{2\tau S}\ave{\mathcal{Q}\mathcal{Q}}_{0} & \displaystyle \lim_{\tau \to \infty} \frac{1}{2\tau}\ave{\mathcal{Q}\mathcal{X}}_{0}\\[8pt]
  \displaystyle \lim_{\tau \to \infty} \frac{1}{2\tau}\ave{\mathcal{X}\mathcal{Q}}_{0} & \displaystyle \lim_{\tau \to \infty} \frac{S}{2\tau}\ave{\mathcal{X}\mathcal{X}}_{0}
 \end{pmatrix}.
 \label{eq:Onsager_coefficient_heat_expression}
\end{equation}

We note that  (\ref{eq:LRformula_K}) and (\ref{eq:LRformula_V}) can be derived by the symmetry of the generating function (\ref{eq:symmetry_generating_func}).
By considering
\begin{equation}
 \frac{\partial G(h_1,h_2)}{\partial h_1}\bigg\vert_{h_1=h_2=0} = \frac{\partial G( \Delta \beta -h_1, \Delta \beta pS-h_2)}{\partial h_1}\bigg\vert_{h_1=h_2=0},
\end{equation}
we obtain
\begin{equation}
 \lim_{\tau \to \infty} \ave{\frac{\mathcal{K}}{\tau S}}_{\Delta} = \lim_{\tau \to \infty} -\frac{1}{\tau S}\frac{\ave{\mathcal{K}\mathrm{e}^{-\Delta \beta \mathcal{Q}}}_{\Delta}}{\ave{\mathrm{e}^{-\Delta \beta \mathcal{Q}}}_{\Delta}}.
\end{equation}
By using (\ref{eq:Integral_FT}), this equation is the same as (\ref{eq:pre_LRformula_K}).
Similarly, by considering
\begin{equation}
 \frac{\partial G(h_1,h_2)}{\partial h_2}\bigg\vert_{h_1=h_2=0} = \frac{\partial G( \Delta \beta -h_1, \Delta \beta pS-h_2)}{\partial h_2}\bigg\vert_{h_1=h_2=0},
\end{equation}
we obtain the same equation as (\ref{eq:LRformula_V}).


Moreover, the McLennan ensemble (\ref{eq:st_dist_LRR}) provides another expression of (\ref{eq:LRformula_V}).
We denote by $\ave{\ }_{\mathrm{st}}$ and $\ave{\ }_{\mathrm{eq}}$ the ensemble averages defined by $P_{\mathrm{st}}$ and $P_{\mathrm{eq}}$, respectively.
Then, by using (\ref{eq:st_dist_LRR}) and $\ave{V}_{\mathrm{eq}}=0$, we obtain
\begin{align}
 \ave{V}_{\mathrm{st}} &=\Delta \beta\ave{V\overline{\mathcal{Q}}}_{\mathrm{eq}} +O(\Delta^2)
 \notag
 \\[5pt]
 & =\Delta \beta \ave{V\left[ \overline{\mathcal{K}} + pS\overline{\mathcal{X}}\right] }_{\mathrm{eq}} + O(\Delta^2),
\label{eq:LRformula_VVV}
\end{align}
which is equivalent to (\ref{eq:LRformula_V}).
Thus, $\widetilde{\mathrm{L}}_{21}$ is also expressed as
\begin{equation}
 \widetilde{\mathrm{L}}_{21} = \ave{V \overline{\mathcal{Q}}}_{\mathrm{eq}}.
\end{equation}
By evaluating the dynamics of $V(t)$ and $\mathcal{K}(\hat{V})$ in equilibrium, one can calculate the Onsager coefficients.

\subsection{Calculation of $\widetilde{\mathrm{L}}_{21}$} \label{sec:Onsager_coefficient}


First, by using perturbation expansion in the small parameter $\epsilon =\sqrt{m/M}$, we can derive the Fokker--Planck equation from (\ref{eq:mastereq_lambda}), and we obtain the time evolution equation of $V(t)$.
In this subsection, we consider the case where $\beta_{\mathrm{L}}=\beta_{\mathrm{R}}$. 
By using a test function, $\Phi (V)$, and the fact that $2\epsilon^2/(1+\epsilon^2) \ll 1$, (\ref{eq:mastereq_lambda}) leads to
\begin{align}
 \int \mathrm{d}V\; \Phi (V) \frac{\partial P(V,t)}{\partial t} &= \int \mathrm{d}v \int \mathrm{d}V''\; \Phi (V)\lambda (v,V'') P(V'',t)
 \notag
 \\[5pt]
 & \qquad - \int \mathrm{d}V\; \Phi (V) \kappa (V) P(V,t)
 \notag
 \\[5pt]
 &= \int \mathrm{d}v \int \mathrm{d}V''\; \sum_{i=1}^{\infty} \frac{1}{i!}\left( \frac{I(v,V'')}{M}\right)^{i} \frac{\partial^{i} \Phi (V'')}{\partial V''^{i}}
 \notag
 \\[5pt]
 &\qquad \times \lambda (v,V'') P(V'',t)
 \notag
 \\[5pt]
 &= \int \mathrm{d}V''\; \Phi (V'') \sum_{i=1}^{\infty} \frac{(-1)^{i}}{i!}\times 
 \notag
 \\[5pt]
 &\qquad \frac{\partial^{i}}{\partial V''^{i}}\left[ \int \mathrm{d}v \left( \frac{I(v,V'')}{M}\right)^{i} \lambda (v,V'') P(V'',t) \right],
\end{align}
where we have used
\begin{equation}
 V=V''+\frac{I(v,V'')}{M}.
\end{equation}
Thus, we can rewrite (\ref{eq:mastereq_lambda}) as the following formal series in powers of $2\epsilon^2/(1+\epsilon^2)$ \cite{Gruber-Lesne}:
\begin{align}
 \frac{\partial P(V,t)}{\partial t} &= \sum_{i=1}^{\infty} \frac{(-1)^{i}}{i!}\frac{\partial^{i}}{\partial V^{i}}\left[ \int \mathrm{d}v\; \left( \frac{I(v,V)}{M}\right)^{i}  \lambda (v,V) P(V,t)\right]
 \notag
 \\[5pt]
 &=\sum_{i=1}^{\infty} \frac{(-1)^{i}}{i!}\left( \frac{2\epsilon^2}{1+\epsilon^2}\right)^{i}\frac{\partial^{i}}{\partial V^{i}}\left[ \int \mathrm{d}v\; (v-V)^{i}  \lambda (v,V) P(V,t)\right].
 \label{eq:pre_FP_eq}
\end{align}
By considering $f_{\mathrm{eq}}^{\mathrm{L}}(v)=f_{\mathrm{eq}}^{\mathrm{R}}(v)=\epsilon \sqrt{\beta M/(2\pi)}\e^{-\epsilon^2 \beta Mv^2/2}$, we obtain the following formula:
\begin{align}
 \int_{V}^{\infty} dv\; v^l f_{\mathrm{eq}}^{\mathrm{L}}(v) &= \int_{0}^{\infty} dv\; v^l f_{\mathrm{eq}}^{\mathrm{L}}(v) + O(\epsilon)
 \notag
 \\[5pt]
 & = \epsilon^{-l}\frac{1}{2\sqrt{\pi}} \left( \frac{2}{\beta M}\right)^{l/2} \mathrm{\Gamma} \left( \frac{l+1}{2}\right) +O(\epsilon),
 \label{eq:useful_formula_L}
 \\[5pt]
  \int_{-\infty}^{V} dv\; v^l f_{\mathrm{eq}}^{\mathrm{R}}(v) &= \int_{-\infty}^{0} dv\; v^l f_{\mathrm{eq}}^{\mathrm{R}}(v) + O(\epsilon)
 \notag
 \\[5pt]
 & = \epsilon^{-l}\frac{(-1)^{l}}{2\sqrt{\pi}} \left( \frac{2}{\beta M}\right)^{l/2} \mathrm{\Gamma} \left( \frac{l+1}{2}\right) +O(\epsilon),
 \label{eq:useful_formula_R}
\end{align}
where $l$ is a non-negative integer.
Then, using (\ref{eq:lambda_L}), (\ref{eq:lambda_R}), (\ref{eq:pre_FP_eq}), (\ref{eq:useful_formula_L}), and (\ref{eq:useful_formula_R}), we obtain the Fokker--Planck equation up to order $\epsilon^2$ as
\begin{equation}
 \frac{\partial P(V,t)}{\partial t} = -\frac{\partial}{\partial V}\left[ \left( -\frac{\gamma V}{M}\right) P(V,t)\right]+\frac{\gamma}{\beta M^2}\frac{\partial^2 P(V,t)}{\partial V^2},
  \label{eq:FP_eq}
\end{equation}
with
\begin{align}
 \gamma &\equiv \epsilon (n_{\mathrm{L}}+n_{\mathrm{R}}) S\sqrt{\frac{8M}{\pi \beta}}
 \notag
 \\
 &= 4\epsilon pS\sqrt{\frac{2\beta M}{\pi}},
\label{eq:gamma}
\end{align}
where $\gamma$ is interpreted as a friction constant.
It should be noted that the friction effect originates from the change in the collision rate due to the motion of the wall.
(\ref{eq:FP_eq}) leads to the following time evolution equation of $V(t)$ \cite{Gardiner}:
\begin{equation}
 M\frac{d V(t)}{dt} = -\gamma V(t) + \sqrt{\frac{2\gamma}{\beta}} \xi(t),
 \label{eq:V_evo}
\end{equation}
where $\xi$ is Gaussian white noise with $\ave{\xi(t)\xi(t')}_{0}=\delta(t-t')$.
By solving (\ref{eq:V_evo}), we obtain
\begin{equation}
 V(t) = V(0) e^{-\frac{\gamma}{M}t} + \sqrt{\frac{2\gamma }{\beta M^2}} \int_{0}^{t} ds\; e^{-(t-s)\frac{\gamma}{M}} \xi(s).
  \label{eq:sol_V}
\end{equation}


Next, we derive the time evolution equation of $\mathcal{K}(\hat{V})$.
We define the energy and heat flux from the gas particles on the left side to the wall by
\begin{equation}
 k_{\mathrm{L}}(\tau ; \hat{V}) \equiv \frac{1}{S}\frac{\mathrm{d}\mathcal{K}_{\mathrm{L}}(\hat{V})}{\mathrm{d}\tau},
\end{equation}
and
\begin{align}
 q_{\mathrm{L}}(\tau ; \hat{V}) &\equiv \frac{1}{S}\frac{\mathrm{d}\mathcal{Q}_{\mathrm{L}}(\hat{V})}{\mathrm{d}\tau}
 \notag
 \\
 &= k_{\mathrm{L}}(\tau ;\hat{V}) - pV(\tau),
  \label{eq:def_J_L}
\end{align}
respectively.
$k_{\mathrm{R}}(\tau ;\hat{V})$ and $q_{\mathrm{R}}(\tau ;\hat{V})$ are defined as well.
We denote the energy and heat flux from right to left by
\begin{equation}
 k(\tau ;\hat{V}) \equiv \frac{k_{\mathrm{R}}(\tau ;\hat{V})-k_{\mathrm{L}}(\tau ;\hat{V})}{2},
\end{equation}
and
\begin{equation}
 q(\tau ;\hat{V}) \equiv \frac{q_{\mathrm{R}}(\tau ;\hat{V})-q_{\mathrm{L}}(\tau ;\hat{V})}{2},
\end{equation}
respectively.
We also denote by $K(v,V)$ the change in the kinetic energy of the wall for the elastic collision of the wall of velocity $V$ with a gas particle of velocity $v$, which is given by
\begin{equation}
 K(v,V) = -\frac{2\epsilon^2}{(1+\epsilon^2)^2}M(V-v)(V+\epsilon^2 v).
\end{equation}
By considering the time evolution equation of the joint distribution function for $V$ and $\mathcal{K}_{\mathrm{L}}$, we obtain in the same way as in deriving (\ref{eq:V_evo})
\begin{equation}
k_{\mathrm{L}}(t) \simeq \int dv\; \lambda_{\mathrm{L}}(v,V(t)) K(v,V(t)) / S,
\label{eq:KL_evo_pre}
\end{equation}
where $\simeq$ means that we ignore the fluctuation terms, which vanish when we take the average value of them.
It should be noted that when we ignore the fluctuation terms, $k_{\mathrm{L}}(t)$ is determined uniquely by $V(t)$.
By using (\ref{eq:useful_formula_L}), we can rewrite (\ref{eq:KL_evo_pre}) up to order $\epsilon^2$ as 
\begin{equation}
 k_{\mathrm{L}}(t) \simeq \left( 1-4\epsilon^2 +\epsilon^2 \beta MV(t)^2\right) pV(t) - \frac{\gamma}{MS}\left( \frac{1}{2}MV(t)^2 -\frac{1}{2\beta}\right) .
\label{eq:k_L_evo}
\end{equation}
Similarly, we obtain up to order $\epsilon^2$
\begin{equation}
  k_{\mathrm{R}}(t) \simeq -\left( 1-4\epsilon^2 +\epsilon^2 \beta MV(t)^2\right) pV(t) - \frac{\gamma}{MS}\left( \frac{1}{2}MV(t)^2 -\frac{1}{2\beta}\right) .
\label{eq:k_R_evo}
\end{equation}
Thus, (\ref{eq:k_L_evo}) and (\ref{eq:k_R_evo}) lead to
\begin{align}
 k(t) &= \frac{k_{\mathrm{R}}(t)-k_{\mathrm{L}}(t)}{2}
 \notag
 \\[3pt]
 &\simeq -\left( 1-4\epsilon^2 +\epsilon^2 \beta MV(t)^2\right) pV(t) + O(\epsilon^3),
  \label{eq:k_evo}
\end{align}
and
\begin{align}
 q(t) &= k(t) + pV(t)
 \notag
 \\
 &\simeq \epsilon^2 \left( 4 - \beta MV(t)^2\right) pV(t) + O(\epsilon^3).
  \label{eq:q_evo}
\end{align}

Finally, we calculate $\widetilde{\mathrm{L}}_{21}$ explicitly.
To lowest order in $\epsilon$, (\ref{eq:sol_V}) and (\ref{eq:q_evo}) lead to
\begin{align}
 \overline{\mathcal{X}}(V) &= \lim_{\tau \to \infty} \ave{\int_0^{\tau} dt\; V(t) \vert_{V(0)=V}}_{0}
 \notag
 \\
 &=\frac{MV}{\gamma},
\label{eq:Delta_X}
\end{align}
and 
\begin{align}
  \overline{\mathcal{Q}}(V) &= \lim_{\tau \to \infty} \ave{\int_0^{\tau} dt\; Sq(t) \vert_{V(0)=V}}_{0}
 \notag
 \\
 &=2\epsilon^2 \frac{pSMV}{\gamma}-\epsilon^2 \frac{pS\beta M^2 V^3}{3\gamma}.
\label{eq:Delta_Q}
\end{align}
Using (\ref{eq:Delta_Q}), $\ave{V^2}_{\mathrm{eq}}=1/(\beta M)$, and $ \ave{V^4}_{\mathrm{eq}}=3/(\beta^2 M^2)$, we obtain to lowest order in $\epsilon$
\begin{align}
 \widetilde{\mathrm{L}}_{21} &= \ave{V\overline{\mathcal{Q}}}_{\mathrm{eq}}
 \notag
 \\
 &=\frac{\epsilon^2 pS}{\beta\gamma}.
 \label{eq:Onsager_coefficient_value}
\end{align}
Using (\ref{eq:Onsager_heat}), (\ref{eq:gamma}), and (\ref{eq:Onsager_coefficient_value}), we obtain up to order $\epsilon$
\begin{align}
 J_{V} &= \widetilde{\mathrm{L}}_{21} \Delta \beta
 \notag
 \\[3pt]
 &= \Delta \epsilon^2 \frac{pS}{\gamma}
 \notag
 \\[3pt]
 &=\frac{\Delta \epsilon}{4} \sqrt{\frac{\pi}{2\beta M}}.
 \label{eq:J_V}
\end{align}
Within the linear response regime, this result is consistent with the previous study \cite{Gruber-Piasecki}.

\section{Concluding Remarks} \label{sec:CR}


In this paper, we elucidate the energetics of the adiabatic piston problem on the basis of the local detailed balance condition.
Owing to the condition, we can decompose the energy transferred from each gas to the wall into the work and the heat transferred.
In the course of the calculation of the condition, we find that the difference of the escape rates, $\kappa(V_{i})-\kappa(V_{i}^{*})$, contributes to the entropy production.
In addition, by using the condition, we obtain the linear response formula for the steady velocity of the wall and steady energy flux through the wall.
By using perturbation expansion in the small parameter $\epsilon\equiv\sqrt{m/M}$, we derive the steady velocity up to order $\epsilon$.
It should be noted that we can derive the local detailed balance condition for the more general case where the wall consists of many atoms \cite{Itami-Sasa}.


The adiabatic piston problem will be important in future studies.
First, since the wall moves despite the same pressure on both sides, this phenomenon cannot be described by the standard hydrodynamic equations, and thus this problem will provide a good example for studying the hydrodynamic equations more deeply.
In particular, the determination of the boundary condition of the fluctuating hydrodynamic equations may be directly related to the description of the observed phenomenon.
Second, since this problem is the simplest example of two interacting systems, one may obtain a mechanical representation of the information exchange process in two interacting stochastic systems \cite{Ito-Sagawa,Sagawa-Ueda2012}.
As seen in these two examples, one can deepen the understanding of nonequilibrium statistical mechanics by developing the analysis of the adiabatic piston problem.


Before ending the paper, we discuss the force from the bath $\widetilde{F}_{\mathrm{L}}$, which plays an essential role on the phenomenon.
The simplest model of $\widetilde{F}_{\mathrm{L}}$ that yields the $T$-$p$ ensemble of the system in equilibrium is give by a Langevin force \cite{SekimotoJPSJ,SekimotoBOOK}
\begin{equation}
\widetilde{F}_{\mathrm{L}}=pS -\gamma_{\mathrm{L}} V +\sqrt{2 T_{\mathrm L} \gamma_{\mathrm{L}}}\xi_{\mathrm{L}},
\label{force:Lan}
\end{equation} 
where $\xi_{\mathrm{L}}$ is Gaussian white noise with unit variance \cite{Gardiner} and $\gamma_{\mathrm{L}}$ a friction constant.
$\widetilde{F}_{\mathrm{R}}$ is similarly defined.
In this case, we can write the equation of motion of the wall as
\begin{align}
 M\frac{\mathrm{d}V}{\mathrm{d}t}&=\widetilde{F}_{\mathrm L} +\widetilde{F}_{\mathrm R}
 \notag
 \\[3pt]
 & = -(\gamma_{\mathrm{L}}+\gamma_{\mathrm{R}}) V + \sqrt{2 T_{\mathrm L} \gamma_{\mathrm{L}}}\xi_{\mathrm{L}} + \sqrt{2 T_{\mathrm R} \gamma_{\mathrm{R}}}\xi_{\mathrm{R}}.
\end{align}
Then, by considering $\ave{\mathrm{d}V/\mathrm{d}t}_{\mathrm{st}}=\ave{\xi_{\mathrm{L}}}_{\mathrm{st}}=\ave{\xi_{\mathrm{R}}}_{\mathrm{st}}=0$, it is easily confirmed that
\begin{equation}
 \ave{V}_{\mathrm{st}}=0.
\end{equation}
That is, the steady state velocity depends on the type of stochastic force.
This result raises the question what is the condition of proper description of the baths.
To understand the proper bath model in nonequilibrium is of great importance.

\begin{acknowledgements}
The authors thank K. Sekimoto, R. Kawai, T. Sagawa, and T.G. Sano for useful discussions.
The present study was supported by KAKENHI No. 22340109 and No. 25103002, and by the JSPS Core-to-Core program ``Non-equilibrium dynamics of soft-matter and information.''
\end{acknowledgements}

\section*{Appendix: Local Detailed Balance Condition for Hamiltonian Systems}

\begin{figure}
\centering
\includegraphics[scale=0.45]{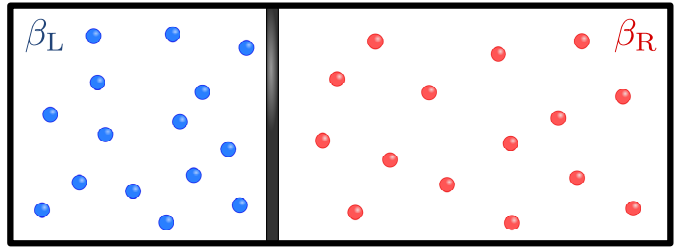}
\caption{(color online) Schematic illustration of our total system, which consists of a wall and two heat baths. The wall consists of only one degree of freedom and separates the left heat bath from the right heat bath. The left and right heat bath consist of $N_{\mathrm{L}}$ and $N_{\mathrm{R}}$ particles, and are initially prepared at different inverse temperatures, $\beta_{\mathrm{L}}$ and $\beta_{\mathrm{R}}$, respectively.}
\label{appendix_fig1}
\end{figure}

We provide the microscopic description of our model.
The model is illustrated in Fig.~\ref{appendix_fig1}.
The system consists of a wall and two heat baths.
We assume that the wall of area $S$ consists of only one degree of freedom and separates the left heat bath from the right heat bath.
The wall moves freely along a long tube and we take $x$-axis as the direction of movement of the wall.
The position and momentum of the wall are denoted as $\gamma = (X, P)$.
We also assume that the left and right heat bath consist of $N_{\mathrm{L}}$ and $N_{\mathrm{R}}$ particles, respectively.
A collection of the positions and momenta of $N_{\mathrm{L}}$ particles in the left heat bath is denoted as $\Gamma^{\mathrm{L}} = (\bm{r}_{1}^{\mathrm{L}},\dots ,\bm{r}_{N_{\mathrm{L}}}^{\mathrm{L}},\bm{p}_{1}^{\mathrm{L}},\dots ,\bm{p}_{N_{\mathrm{L}}}^{\mathrm{L}})$, and that of $N_{\mathrm{R}}$ particles in the right heat bath is similarly denoted as $\Gamma^{\mathrm{R}}$.
Then, the microscopic state of the total system is expressed as a point of the phase space $\Gamma \equiv (\gamma ,\Gamma^{\mathrm{L}}, \Gamma^{\mathrm{R}})$.
For any state $\Gamma$, we denote by $\Gamma^{*}$ its time reversal, namely, the state obtained by reversing all the momenta, and denote the time reversal of $\bm{p}_{i}$ as $\bm{p}_{i}^{*} = -\bm{p}_{i}$.
We assume that all interactions are short-range, and ignore the interaction between the left and right heat bath particles for simplicity.
We denote by $\mathcal{H}(\gamma)$, $H_{\mathrm{L}}(\Gamma^{\mathrm{L}}; \gamma)$, and $H_{\mathrm{R}}(\Gamma^{\mathrm{R}}; \gamma)$ the Hamiltonian of the wall, the left heat bath, and the right heat bath, respectively, where $H_{\mathrm{L}}(\Gamma^{\mathrm{L}}; \gamma)$ and $H_{\mathrm{R}}(\Gamma^{\mathrm{R}}; \gamma)$ include the interaction potential between the particles in each heat bath and the wall.
Then, the time evolution of $\Gamma$ is determined by the following total Hamiltonian:
\begin{equation}
 H(\Gamma) \equiv \mathcal{H}(\gamma) +H_{\mathrm{L}}(\Gamma^{\mathrm{L}}; \gamma) +H_{\mathrm{R}}(\Gamma^{\mathrm{R}}; \gamma).
  \label{eq:total_Hamiltonian_app}
\end{equation}
We denote by $\Gamma_{t}$ the solution of the Hamiltonian equations at time $t$ for any initial state $\Gamma$.
We assume that the total Hamiltonian satisfies the time-reversal symmetry:
\begin{equation}
 H(\Gamma)=H(\Gamma^{*}).
  \label{eq:time_reversal_symmetry_app}
\end{equation}
In this setup, we obtain Liouville's theorem:
\begin{equation}
 \left\vert \frac{\partial \Gamma_{t}}{\partial \Gamma} \right\vert =1,
  \label{eq:Liouville_app}
\end{equation}
and the law of conservation of energy:
\begin{equation}
 H(\Gamma_{t})=H(\Gamma).
  \label{eq:conservation_energy_app}
\end{equation}
In the following, we consider the time evolution in the time interval $[0,\tau]$.

Next, we determine initial conditions.
We fix the initial state of the wall as $\gamma = \gamma_{\mathrm{i}}$ and the final state of the wall as $\gamma_{\tau} = \gamma_{\mathrm{f}}$.
We initially prepare the left and right heat baths at different inverse temperatures, $\beta_{\mathrm{L}}$ and $\beta_{\mathrm{R}}$, respectively.
Thus, the initial states of the heat baths are sampled according to the probability density
\begin{equation}
 P_{\mathrm{B}}(\Gamma^{\mathrm{L}}, \Gamma^{\mathrm{R}}; \gamma) = \mathrm{e}^{\beta_{\mathrm{L}}[ F_{\mathrm{L}}(\gamma) - H_{\mathrm{L}}(\Gamma^{\mathrm{L}}; \gamma)] + \beta_{\mathrm{R}}[ F_{\mathrm{R}}(\gamma) - H_{\mathrm{R}}(\Gamma^{\mathrm{R}}; \gamma)]},
  \label{eq:P_B_app}
\end{equation}
with
\begin{equation}
 F_{\mathrm{L}}(\gamma) = -\frac{1}{\beta_{\mathrm{L}}} \log \left[ \int \mathrm{d}\Gamma^{\mathrm{L}}\; \mathrm{e}^{-\beta_{\mathrm{L}} H_{\mathrm{L}}(\Gamma^{\mathrm{L}}; \gamma)}\right] ,
  \label{eq:F_L_app}
\end{equation}
\begin{equation}
 F_{\mathrm{R}}(\gamma) = -\frac{1}{\beta_{\mathrm{R}}} \log \left[ \int \mathrm{d}\Gamma^{\mathrm{R}}\; \mathrm{e}^{-\beta_{\mathrm{R}} H_{\mathrm{R}}(\Gamma^{\mathrm{R}}; \gamma)}\right] ,
  \label{eq:F_R_app}
\end{equation}
where $F_{\mathrm{L}}(\gamma)$ and $F_{\mathrm{R}}(\gamma)$ represent the Helmholtz free energies of the left and right heat baths, respectively.
By using the Helmholtz free energy, we can define the pressure of the left heat bath to the wall by
\begin{align}
 p_{\mathrm{L}}(\gamma) &\equiv -\frac{\partial F_{\mathrm{L}}(\gamma )}{\partial (SX)}
 \notag
 \\[3pt]
 &= \int \mathrm{d}\Gamma^{\mathrm{L}}\; \mathrm{e}^{\beta_{\mathrm{L}} [F_{\mathrm{L}}(\gamma)-H_{\mathrm{L}}(\Gamma^{\mathrm{L}}; \gamma)]} \left[ -\frac{\partial H_{\mathrm{L}}(\Gamma^{\mathrm{L}}; \gamma)}{\partial (SX)}\right] .
  \label{eq:pL_app}
\end{align}
The pressure of the right heat bath to the wall is similarly defined by
\begin{equation}
 p_{\mathrm{R}}(\gamma )\equiv \frac{\partial F_{\mathrm{R}}(\gamma)}{\partial (SX)}.
  \label{eq:pR_app}
\end{equation}
It should be noted that the direction of the pressure force to the wall on the left side is opposite to that on the right side.
In this setup, we can write the probability density for paths of the wall starting at $\gamma_{\mathrm{i}}$ and ending at $\gamma_{\mathrm{f}}$ as
\begin{equation}
 \mathcal{W}(\gamma_{\mathrm{i}}\to \gamma_{\mathrm{f}}) = \int \mathrm{d}\Gamma \; P_{\mathrm{B}}(\Gamma^{\mathrm{L}}, \Gamma^{\mathrm{R}}; \gamma) \delta (\gamma - \gamma_{\mathrm{i}}) \delta (\gamma_{\tau} - \gamma_{\mathrm{f}}),
  \label{eq:W_app}
\end{equation}
which satisfies
\begin{equation}
 \int \mathrm{d}\gamma_{\mathrm{f}}\; \mathcal{W}(\gamma_{\mathrm{i}}\to \gamma_{\mathrm{f}}) = 1.
\end{equation}
For any physical quantity $A(\Gamma)$, we define its time reversal by
\begin{equation}
 A^{\dagger} (\Gamma) \equiv A(\Gamma_{\tau}^{*}),
\end{equation}
and the ensemble average of $A(\Gamma)$ with the initial and final conditions of the wall $\gamma_{\mathrm{i}}$ and $\gamma_{\mathrm{f}}$, respectively, by
\begin{equation}
 \ave{A}_{\gamma_{\mathrm{i}}\to \gamma_{\mathrm{f}}} \equiv \frac{\int \mathrm{d}\Gamma \; P_{\mathrm{B}}(\Gamma^{\mathrm{L}}, \Gamma^{\mathrm{R}}; \gamma) \delta (\gamma - \gamma_{\mathrm{i}}) \delta (\gamma_{\tau} - \gamma_{\mathrm{f}}) A(\Gamma)}{\mathcal{W}(\gamma_{\mathrm{i}}\to \gamma_{\mathrm{f}})}.
\end{equation}

Finally, we derive the local detailed balance condition.
We define the decrease in the internal energy of the left and right heat bath by
\begin{equation}
 \begin{cases}
  U_{\mathrm{L}}(\Gamma) \equiv H_{\mathrm{L}}(\Gamma^{\mathrm{L}}; \gamma) -H_{\mathrm{L}}(\Gamma^{\mathrm{L}}_{\tau}; \gamma_{\tau}),
  \\[5pt]
  U_{\mathrm{R}}(\Gamma) \equiv H_{\mathrm{R}}(\Gamma^{\mathrm{R}}; \gamma) -H_{\mathrm{R}}(\Gamma^{\mathrm{R}}_{\tau}; \gamma_{\tau}).
  \end{cases}
  \label{eq:internal_energy_app}
\end{equation}
By using (\ref{eq:time_reversal_symmetry_app}) and (\ref{eq:P_B_app}), we obtain
\begin{equation}
 \frac{P_{\mathrm{B}}(\Gamma^{\mathrm{L}}, \Gamma^{\mathrm{R}}; \gamma)}{P_{\mathrm{B}}(\Gamma^{\mathrm{L}*}_{\tau}, \Gamma^{\mathrm{R}*}_{\tau}; \gamma_{\tau}^{*})} = \mathrm{e}^{\beta_{\mathrm{L}}[F_{\mathrm{L}}(\gamma^{*}) -F_{\mathrm{L}}(\gamma_{\tau}^{*}) - U_{\mathrm{L}}(\Gamma)] +\beta_{\mathrm{R}}[F_{\mathrm{R}}(\gamma^{*}) -F_{\mathrm{R}}(\gamma_{\tau}^{*})-U_{\mathrm{R}}(\Gamma)]}.
\end{equation}
Thus, according to Liouville's theorem (\ref{eq:Liouville_app}), we obtain
\begin{align}
 \mathcal{W}(\gamma_{\mathrm{i}}\to \gamma_{\mathrm{f}}) \ave{A}_{\gamma_{\mathrm{i}}\to \gamma_{\mathrm{f}}} &= \int \mathrm{d}\Gamma\; P_{\mathrm{B}}(\Gamma^{\mathrm{L}}, \Gamma^{\mathrm{R}}; \gamma) \delta (\gamma - \gamma_{\mathrm{i}}) \delta (\gamma_{\tau} - \gamma_{\mathrm{f}}) A(\Gamma)
 \notag
 \\[5pt]
 & = \int \mathrm{d}\Gamma_{\tau}^{*}\; P_{\mathrm{B}}(\Gamma^{\mathrm{L}*}_{\tau}, \Gamma^{\mathrm{R}*}_{\tau}; \gamma_{\tau}^{*}) \delta (\gamma_{\tau}^{*} - \gamma_{\mathrm{f}}^{*}) \delta (\gamma^{*} - \gamma_{\mathrm{i}}^{*})  A^{\dagger}(\Gamma_{\tau}^{*})
 \notag
 \\[5pt]
 & \quad \times \mathrm{e}^{\beta_{\mathrm{L}}[F_{\mathrm{L}}(\gamma^{*}) -F_{\mathrm{L}}(\gamma_{\tau}^{*}) - U_{\mathrm{L}}^{\dagger}(\Gamma_{\tau}^{*})] +\beta_{\mathrm{R}}[F_{\mathrm{R}}(\gamma^{*}) -F_{\mathrm{R}}(\gamma_{\tau}^{*})-U_{\mathrm{R}}^{\dagger}(\Gamma_{\tau}^{*})]}.
 \label{eq:LDB_pre_app}
\end{align}
Here we assume that the displacement of the wall is much shorter than the length of the long tube in the $x$-direction, and therefore that the pressures of the left and right heat bath are kept constant over time.
Furthermore, $F_{\mathrm{L}}$ depends only on $X$.
Then, by using (\ref{eq:pL_app}), we obtain
\begin{equation}
 F_{\mathrm{L}}(\gamma_{\tau}) = F_{\mathrm{L}}(\gamma) - p_{\mathrm{L}}(\gamma) S(X_{\tau}-X).
  \label{eq:F_L_Taylor_app}
\end{equation}
$p_{\mathrm{L}}(\gamma) S(X_{\tau}-X)$ is the work done by the particles in the left heat bath.
Here, with the following definitions
\begin{equation}
 \begin{cases}
  Q_{\mathrm{L}}(\Gamma) \equiv U_{\mathrm{L}}(\Gamma) - p_{\mathrm{L}}(\gamma )S(X_{\tau}-X),
  \\[5pt]
  Q_{\mathrm{R}}(\Gamma) \equiv U_{\mathrm{R}}(\Gamma) + p_{\mathrm{R}}(\gamma )S(X_{\tau}-X),
 \end{cases}
 \label{eq:Heat_def_app}
\end{equation}
(\ref{eq:LDB_pre_app}) leads to
\begin{equation}
 \mathcal{W}(\gamma_{\mathrm{i}}\to \gamma_{\mathrm{f}}) \ave{A}_{\gamma_{\mathrm{i}}\to \gamma_{\mathrm{f}}} = \mathcal{W}(\gamma_{\mathrm{f}}^{*}\to \gamma_{\mathrm{i}}^{*}) \ave{A^{\dagger} \mathrm{e}^{-\beta_{\mathrm{L}}Q_{\mathrm{L}}^{\dagger} -\beta_{\mathrm{R}} Q_{\mathrm{R}}^{\dagger}}}_{\gamma_{\mathrm{f}}^{*}\to \gamma_{\mathrm{i}}^{*}}.
 \label{eq:LDB1_app}
\end{equation}
From the first law of thermodynamics, $Q_{\mathrm{L}}(\Gamma)$ and $Q_{\mathrm{R}}(\Gamma)$ are interpreted as the heat transferred from the left and right heat bath to the wall, respectively.
Now, we define the mean inverse temperature by $\beta = (\beta_{\mathrm{L}}+\beta_{\mathrm{R}})/2$, the degree of nonequilibrium by $\Delta = (\beta_{\mathrm{L}}-\beta_{\mathrm{R}}) / \beta$, and the Helmholtz free energy for the wall by
\begin{equation}
 F \equiv -\frac{1}{\beta}\log \left[ \int \mathrm{d}\gamma\; \mathrm{e}^{-\beta \mathcal{H}(\gamma)}\right].
\end{equation}
By using (\ref{eq:conservation_energy_app}), we obtain
\begin{equation}
 U_{\mathrm{L}}(\Gamma)+U_{\mathrm{R}}(\Gamma)=\mathcal{H}(\gamma_{\tau})-\mathcal{H}(\gamma).
\end{equation}
Thus, by setting $p_{\mathrm{L}}(\gamma_{\mathrm{i}})=p_{\mathrm{R}}(\gamma_{\mathrm{i}})=p$, we can rewrite (\ref{eq:LDB1_app}) as
\begin{equation}
 \widetilde{P}_{\mathrm{eq}}(\gamma_{\mathrm{i}}) \mathcal{W}(\gamma_{\mathrm{i}}\to \gamma_{\mathrm{f}}) \ave{A}_{\gamma_{\mathrm{i}}\to \gamma_{\mathrm{f}}} = \widetilde{P}_{\mathrm{eq}}(\gamma_{\mathrm{f}}^{*}) \mathcal{W}(\gamma_{\mathrm{f}}^{*}\to \gamma_{\mathrm{i}}^{*}) \ave{A^{\dagger} \mathrm{e}^{\Delta \beta \frac{Q_{\mathrm{R}}^{\dagger}-Q_{\mathrm{L}}^{\dagger}}{2}}}_{\gamma_{\mathrm{f}}^{*}\to \gamma_{\mathrm{i}}^{*}},
  \label{eq:LDB2_app}
\end{equation}
with
\begin{equation}
 \widetilde{P}_{\mathrm{eq}}(\gamma) \equiv \mathrm{e}^{\beta [F-\mathcal{H}(\gamma)]}.
\end{equation}
By setting $A=1$, (\ref{eq:LDB1_app}) and (\ref{eq:LDB2_app}) are consistent with the local detailed balance conditions in the main text (\ref{eq:LDB}) and (\ref{eq:LDB2}), respectively.
The form of the heat transferred and the work from each gas to the wall in our stochastic model is also consistent with that in the description of Hamiltonian systems under the assumption that gas regions are so large that thermodynamic quantities of the heat baths are kept constant over observation time.

\end{document}